# Theory of Spatial Optical Solitons in Metallic Nanowire Materials

*Mário G. Silveirinha*

*University of Coimbra, Department of Electrical Engineering – Instituto de Telecomunicações, Portugal, mario.silveirinha@co.it.pt*

**Abstract**

We characterize the spatial optical solitons supported by arrays of metallic nanowires embedded in Kerr-type material. The array of nanowires is described using an effective medium model and is regarded as a continuous medium. It is shown that the conditions necessary for the formation of spatial-solitons are radically different in presence of the nanowires, and in particular within the effective medium model spatial-solitons are allowed in the nanowire material only in case the host material is a "self-defocusing" material. It is proven that the characteristic soliton beamwidth is related to the degree of hyperbolicity of the isofrequency surfaces of the photonic states, and that a sufficiently strong electric field amplitude may enable subwavelength solitary waves.

# I. Introduction

Light propagation in nanowire arrays has been a topic of intense research in recent years due to the opportunity provided by such structures to manipulate electromagnetic radiation in a spatial scale much smaller than the wavelength [1]-[4]. More recently, the impact of nonlinear effects in the light propagation in metallic nanowire arrays and other closely related nanostructures has been investigated [5, 6, 9], and it has been predicted that such "waveguide arrays" may support subwavelength spatial-solitons [7-8] when the metallic inclusions are embedded in a nonlinear Kerr-type dielectric host. This may enable achieving the subwavelength confinement of light guided by metallic nanostructures, and help reducing the characteristic size of photonic devices. Several studies have demonstrated that spatial-solitons (also known as *plasmonic lattice solitons*) in nanowire arrays are stable when the nanowires are embedded in a self-defocusing medium, and a variety of vortex and multipole have been studied in Refs. [10, 11]. The analysis of these works is based on the coupled-mode theory, which provides a limited physical insight of the problem. Recently [12], building on the efforts of several earlier works [13, 14, 15], we developed an effective medium model that enables regarding the array of metallic nanowires embedded in a Kerr-type material as a continuous medium characterized by some nonlinear effective parameters. It was shown that the nanowire array may behave as a hyperbolic (indefinite) uniaxial medium (such that the signs of the components of the permittivity tensor differ), and that a weak nonlinearity enables to control the degree of hyperbolicity of the dispersion of the photonic states. In particular, it was predicted that the nonlinear effects may enhance the negative refraction of light at an interface between air and the nanowire array. In this work, we apply the effective medium model to the study of spatial-solitons in nanowire arrays embedded in a nonlinear host. Based on the effective medium framework, we explain the mechanism of formation of the spatial-solitons. We prove that the interplay between the nonlinearity of the host and the waveguiding properties of the plasmonic inclusions, results in an effective medium response such that if the host dielectric (i.e. the nonlinear component of the composite material) is a self-*defocusing* medium then the array of nanowires behaves as a self-*focusing* medium. We develop an analytical theory to characterize the spatial solitons, and present a parametric numerical study of the properties of the spatial solitons in simple two-dimensional scenarios, including the effect of loss.

## II. Trapped States in Nonlinear Hyperbolic Media

In this section, we discuss from a qualitative point of view the physical requirements for the formation of trapped states in arrays of metallic nanowires embedded in a nonlinear host medium with a Kerr-type nonlinearity. The array of nanowires is treated as an effective medium, and to a rough approximation it may be regarded as a hyperbolic medium. This is so because the isofrequency diagrams of the photonic states associated with the so-called quasi-transverse electromagnetic (TEM) are hyperbolic contours [14, 16]. As discussed in detail in our previous work [12], the shape of the isofrequency diagrams depends on the nonlinear effects, and thus on the intensity of the electromagnetic field.

The geometry of the nanowire array is illustrated in Fig. 1. It is assumed that the metallic nanowires are cylindrical and oriented along the *z*-direction, and have radius $r_w$. The metal electrical permittivity is $\varepsilon_m$. The period of the array is *a* and the nanowires are embedded in a host medium with a Kerr-type nonlinearity such that $\varepsilon_h = \varepsilon_h^0 (1 + \delta\varepsilon)$ with $\delta\varepsilon = \alpha \mathbf{e}^* \cdot \mathbf{e}$, being $\mathbf{e}$ the microscopic electric field (i.e. before any averaging on the scale of the period of the array) and $\alpha = 3\chi^{(3)} / \varepsilon_{h,r}^0$ is a constant that determines the strength of the nonlinear effects.

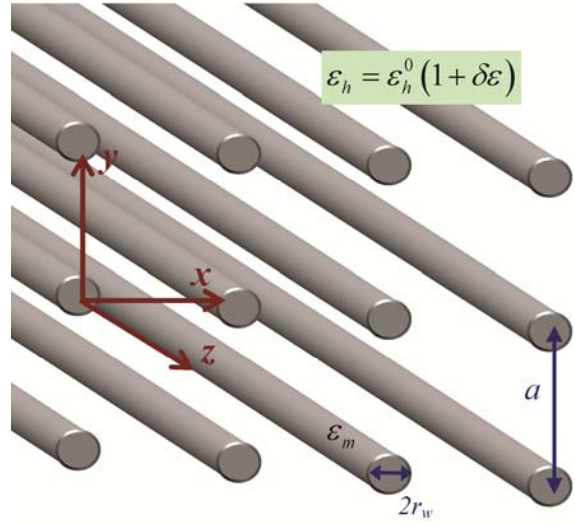

Fig. 1. (Color online) Illustration of a periodic metallic nanowire array embedded in a Kerr-type nonlinear host material.

The mechanism of formation of trapped states in conventional self-focusing media is well-known, and can be understood from the dynamics of the isofrequency contours of the photonic states as a function of the nonlinear effects. Let us suppose that the direction of propagation is along $z$, and that $k_{z0}$ is the wave number which determines the variation in space of the trapped state along the direction of propagation. For a fixed frequency, a trapped state associated with certain $k_{z0}$ is allowed if for a sufficiently strong field amplitude the medium supports a photonic state with $k_z = k_{z0}$ (such that a generic photonic state is characterized by the wave vector $\mathbf{k} = \mathbf{k}_t + k_z \hat{\mathbf{z}}$, being $\mathbf{k}_t = k_x \hat{\mathbf{x}} + k_y \hat{\mathbf{y}}$) and simultaneously for weak field amplitudes there are no allowed photonic states with $k_z = k_{z0}$. In such circumstances, the radiation can become trapped in case of sufficiently strong field amplitudes because it cannot be coupled to photonic states lying in the regions of weak field amplitude. Thus the trapping mechanism is related to the total-internal reflection due to the self-induced positive refractive index change.

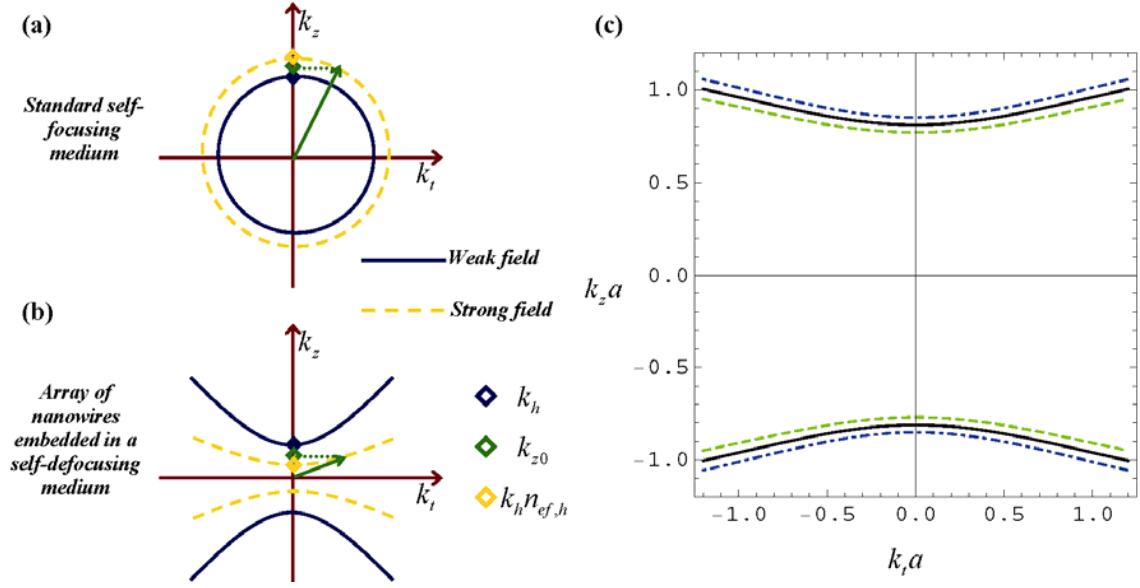

Fig. 2. (Color online) Illustration of the mechanism that enables the formation of spatial solitons. (a) Dispersion of photonic states in a standard self-focusing medium for the cases of weak (solid blue line) and strong (dashed yellow line) field amplitudes. (b) Similar to (a) but for an array of nanowires embedded in a self-defocusing medium. (c) Dispersion of the photonic states associated with the extraordinary (q-TEM) wave at $\lambda_0$=1550nm in a nanowire material formed by silver wires in a dielectric background with $\varepsilon_h^0 = 1.0\varepsilon_0$, $r_w = 20nm$, $r_w/a = 0.1$. The dispersion of the photonic states is obtained based on the theory of Ref. [12]. Solid (black) lines: $n_{ef,h}^2 = n_w^2/\zeta_w = 1.0$ (linear host material); Dot-dashed (blue) lines: $n_{ef,h}^2 = n_w^2/\zeta_w = 1.0 + 0.1$ (self-focusing host material); Dashed (green) lines: $n_{ef,h}^2 = n_w^2/\zeta_w = 1.0 - 0.1$ (self-defocusing host material). The green arrow represents a photonic state with $k_z = k_{z0}$ such that it only propagates in case of a sufficiently strong field.

These ideas are illustrated in Fig. 2a, which shows the dispersion of the photonic states $k^2 c^2 = \omega^2 n^2$ ($c$ is the speed of light in vacuum) for a conventional self-focusing medium such that index of refraction $n$ grows with the amplitude of the electric field.

In order to study when solitary waves are allowed in a nanowire material, we depict in Fig. 2a the isofrequency contours of the quasi-TEM mode in case of weak fields at 1550 nm for a representative material geometry (solid black line). It is assumed that the nanowires are made of silver. The dispersion of silver is described by the Drude permittivity model with a plasma frequency 2175 THz [20]. As seen, the isofrequency surface of the photonic states resembles a hyperbola, rather different from what happens in a vacuum where it is a spherical surface.

In case the nanowires are embedded in a nonlinear material the isofrequency contours depend on the effective parameters $n_{ef,h}^2$ and $n_w^2/\zeta_w$, whose precise definition is given in Ref. [12] (see also Sect. III). When the nanowires are embedded in a self-focusing (self-defocusing) nonlinear material $n_{ef,h}^2$ and $n_w^2/\zeta_w$ are greater (smaller) than unity and grow (become smaller) as the amplitude of the fields grows. For very weak fields (linear approximation) we have $n_{ef,h}^2 = n_w^2/\zeta_w = 1$. The perturbation in the dispersion of the photonic states caused by the nonlinear effects is also represented in Fig. 2c with dashed green and dot-dashed blue lines for the cases of a self-defocusing and self-focusing host medium, respectively. It can be checked based on the analytical model of Ref. [12] that the point of the hyperbola with $k_t = 0$ is such that $k_z = k_h n_{ef,h}$, being $k_h = \omega\sqrt{\varepsilon_h^0 \mu_0}$ and $\varepsilon_h^0$ the permittivity of the host medium for the case of weak field amplitudes.

Based on the isofrequency contours of Fig. 2c one can readily understand in which circumstances trapped states can be formed in the effective medium. Indeed, taking into account the hyperbolic shape of the isofrequency surfaces, it is seen that the value $k_{z0}$ associated with a trapped state must be such that: *(i)* $k_{z0} > k_h n_{ef,h}$, so that a photonic state is allowed in case of a sufficiently strong field amplitude. *(ii)* $k_{z0} < k_h$, so that no photonic states are supported in case of a weak field. The two conditions can be met simultaneously only if $n_{ef,h} < 1$, i.e. only in case of a self-defocusing host medium. In particular, one concludes that within the effective medium model adopted here no solitary waves are allowed in the nanowire material in case of a self-focusing host medium. It is important to mention that Ref. [6] reported that discrete spatial solitons may be supported by arrays of metallic nanowires embedded in a Kerr self-focusing medium, in apparent contradiction with our theory. However, the two theories can be reconciled by noting that for a self-focusing medium the trapped states found in Ref. [6] are staggered solitons. Such trapped modes have strong spatial field variations on the scale of the period of the nanowire material, and thus cannot be described within an effective medium approach, which is the scope of our study. Figure 2b illustrates the isofrequency contours in a nanowire material with a self-defocusing host, and an allowed value for $k_{z0}$ is marked in the plot.

It is interesting to mention that trapped waves described previously are partly related to gap solitons [17-18], because the dispersion of the photonic states in the nanowire array has a directional bandgap. However, here the spatial soliton formation does not involve Bragg scattering and is rooted on the effective medium response. The emergence of gap solitons in metamaterials has been discussed in [19].

# III. Effective Medium Model for the Characterization of the Spatial Solitons

In order to put the ideas of the previous section in a firm theoretical standing, next we derive the equations that allow calculating the spatial-solitons in the nanowire material. The analysis is based on the effective medium model derived in Ref. [12], which describes the electrodynamics of the wire medium in terms of the state vector $(\mathbf{E}, \mathbf{H}, \varphi_w, I)$, being $(\mathbf{E}, \mathbf{H})$ the macroscopic electromagnetic fields, $I$ the current in the nanowires, and $\varphi_w$ the average quasi-static electric potential drop from a given nanowire to the boundary of the respective unit cell. Both $I$ and $\varphi_w$ are interpolated so that they can be regarded as continuous functions of the position. The state vector $(\mathbf{E}, \mathbf{H}, \varphi_w, I)$ satisfies a nonlinear first order partial-differential system. It is however possible to eliminate $(\mathbf{H}, I)$ in favor of $(\mathbf{E}, \varphi_w)$. This yields the following second order partial-differential system [12, Eq. 36]:

$$\nabla \times \nabla \times \mathbf{E} - k_h^2 n_{ef,h}^2 \mathbf{E} = i\omega\mu_0 \mathbf{j}_{ext} + \frac{\beta_p^2}{\zeta_w}\left(\frac{\partial \varphi_w}{\partial z} - E_z\right)\hat{\mathbf{z}} + \beta_p^2 k_h^2 \tilde{\mathbf{Y}} \varphi_w, \tag{1a}$$

$$\frac{\partial^2 \varphi_w}{\partial z^2} + k_h^2 n_w^2 \varphi_w = -\zeta_w k_h^2 \tilde{\mathbf{Y}} \cdot \mathbf{E}_t + \frac{\partial E_z}{\partial z}. \tag{1b}$$

where $k_h^2 = \omega^2 \varepsilon_h^0 \mu_0$, $\mathbf{E}_t = E_x\hat{\mathbf{x}} + E_y\hat{\mathbf{y}}$ is the transverse electric field component, $\mathbf{j}_{ext}$ is an hypothetical external current density, $\zeta_w = 1 + \frac{Z_w}{-i\omega L}$, $Z_w = -\frac{1}{i\omega\pi r_w^2 (\varepsilon_m - \varepsilon_h^0)}$ is the per unit of length (p.u.l) self-impedance of the nanowires, $L = \frac{\mu_0}{2\pi}\log\left(\frac{a^2}{4r_w(a-r_w)}\right)$ is the

p.u.l. geometrical inductance of the nanowires, and $\beta_p = a^{-1}\sqrt{\mu_0/L}$ is the geometrical component of the effective plasma-wave number of the metamaterial [15].

As a consequence of the nonlinear response of the host medium (described by the parameter $\alpha$), the effective parameters $\tilde{\mathbf{Y}} = \frac{\alpha}{2}\left(\varphi_w \mathbf{E}_t^* + \varphi_w^* \mathbf{E}_t\right)$, $n_{ef,h}^2 = 1 + \alpha\left(\mathbf{E}^* \cdot \mathbf{E} + \beta_p^2 \varphi_w \varphi_w^*\right)$, and $n_w^2 = \zeta_w\left[1 + \alpha\left(\mathbf{E}^* \cdot \mathbf{E} + \tilde{B}\beta_p^2 \varphi_w \varphi_w^*\right)\right]$ are quadratic forms of the state vector. The dimensionless parameter $\tilde{B}$ depends exclusively on the normalized radius of the nanowires, $r_w/a$. An explicit formula for $\tilde{B}$ can be found in Ref. [12]. The system of equations (1) is our starting point for the calculation of the spatial-solitons supported by the nanowire array based on the effective model.

To make further progress it is necessary to simplify somewhat the system (1). We consider two different approaches based on different simplifications of the formulas for $n_{ef,h}^2$, $n_w^2$ and $\tilde{\mathbf{Y}}$. The parameters $n_{ef,h}^2$ and $n_w^2$ depend on $|E_z|^2$, $|E_t|^2$ and $|\varphi_w|^2$, with $|E_t| = \sqrt{|E_x|^2 + |E_y|^2}$. In case of *paraxial* spatial-solitons propagating along the $z$-direction, the light beam is expected to be a quasi-plane wave that interacts relatively weakly with the metallic nanowires, and thus the associated electric field is mainly confined to the *xoy* plane so that $|E_z| \ll |E_t|$. Moreover, because $\varphi_w$ is the quasi-static electric potential created by the electric charge density induced on the metallic wires [12], for paraxial propagation one may also expect that $|\varphi_w/a| \ll |E_t|$. This will be confirmed ahead with numerical simulations. Hence, in the paraxial case we can neglect contributions to the nonlinear dynamics arising from both $E_z$ and $\varphi_w$. Thus, one may assume that:

$$n_{ef,h}^2 \approx \frac{n_w^2}{\zeta_w} \approx 1 + \alpha \mathbf{E}_t^* \cdot \mathbf{E}_t, \quad \text{and} \quad \tilde{\mathbf{Y}} \approx 0. \tag{2}$$

With these simplifications the nonlinear system (1) is considerably simplified, as shown below:

$$\nabla \times \nabla \times \mathbf{E} - k_h^2 n_{ef,h}^2 \mathbf{E} = i\omega\mu_0 \mathbf{j}_{ext} + \frac{\beta_p^2}{\zeta_w}\left(\frac{\partial \varphi_w}{\partial z} - E_z\right)\hat{\mathbf{z}}, \tag{3a}$$

$$\frac{\partial^2 \varphi_w}{\partial z^2} + k_h^2 \zeta_w n_{ef,h}^2 \varphi_w = \frac{\partial E_z}{\partial z}. \tag{3b}$$

Notice that in this approach we retain $E_z$ and $\varphi_w$ because they are necessary to couple the two second-order equations. The double curl operator could be further simplified using the paraxial approximation, but this is not required unless one is interested in studying the effect of perturbations (e.g. loss) in the propagation (see Sect. V).

The other approach that we consider here is based on a less drastic simplification of the effective medium model. As detailed in Appendix A, in this second approach only the contribution from $|E_z|^2$ to the nonlinear dynamics is discarded, so that $n_{ef,h}^2$, $n_w^2$ are given by Eqs. (A1a)-(A1b) and the exact form of $\tilde{\mathbf{Y}}$ is retained. The neglect of $|E_z|^2$ is justified by the fact that even in the non-paraxial case one may expect $|E_z| \ll |E_t|$. Indeed, it is known that in wire media formed by metals with high conductivity the extraordinary wave remains a quasi-transverse electromagnetic (q-TEM) with $|E_z| \ll |E_t|$, even when the fields vary appreciably along the *x* and *y* directions [14, 15, 13]. This is especially accurate in case the radius of the wires is a few times larger than the metal skin depth.

To determine the spatial solitons we solve either Eqs. (3) (paraxial approach) or Eqs. (1) (non-paraxial approach) assuming that the dependence on *z* of $(\mathbf{E}, \varphi_w)$ is of the form $e^{ik_z z}$. For simplicity, we only consider two-dimensional solitary waves with $E_x = 0$ and $\partial_x = 0$. Moreover, for the reasons explained in Appendix A, we are interested in spatial solitons such that $E_y$ and $\varphi_w$ are in phase, and $E_y$ and $E_z$ are in quadrature. Thus, we may look for solutions such that $E_y = \tilde{E}_y(y)e^{ik_z z}$ and $\varphi_w = \tilde{\varphi}_w(y)e^{ik_z z}$ with the envelopes $\tilde{E}_y$ and $\tilde{\varphi}_w$ real valued. It is proven in the Appendix A that within the paraxial approximation [Eqs.(2)-(3)] $E_y$ and $\varphi_w$ satisfy

$$\frac{\partial \varphi_w}{\partial y} = -\frac{k_z^2 - k_h^2 n_{ef,h}^2}{k_h^2 \zeta_w \left(n_{ef,h}^2 + \alpha E_y E_y^*\right) - k_z^2} E_y, \tag{4a}$$

$$\frac{\partial E_y}{\partial y} = -\frac{k_h^2 \zeta_w n_{ef,h}^2 - k_z^2 - \beta_p^2}{n_{ef,h}^2 + 2\alpha E_y E_y^*} n_{ef,h}^2 \varphi_w, \tag{4b}$$

being $n_{ef,h}^2$ given by Eq.(2). In the more general non-paraxial approach [Eqs. (1)-(A1)] the fields $E_y$ and $\varphi_w$ satisfy a similar but more complex first-order nonlinear system [Eq. (A4)]. We restrict our attention to solitons such that $\tilde{E}_y(y)$ is an even function of $y$ and consequently $\tilde{E}_z(y)$ and $\tilde{\varphi}_w(y)$ are odd functions of $y$. Thus, the boundary conditions at $y = 0$ are such that:

$$\left.\varphi_w\right|_{y=0} = 0 \qquad \left.\tilde{E}_y\right|_{y=0} = \tilde{E}_{y0} \tag{5}$$

The value of $\tilde{E}_{y0}$ depends on $k_z$, and is determined iteratively in order to ensure that for $y \to +\infty$ the electromagnetic fields vanish: $E_y, \varphi_w \to 0$.

It is interesting to mention that in case of perfectly electric conducting (PEC) metallic wires the normalized impedance $\zeta_w$ is equal to the unity. In such a case and for a linear host ($n_{ef,h}^2 = n_w^2 / \zeta_w = 1$ and $\tilde{\mathbf{Y}} = 0$) it can be easily checked that when $k_z = k_h$ Eq. (1)

admits a solution with $E_z = 0$ (TEM wave beam). Indeed, for $k_z = k_h$ and $E_z = 0$ Eq. (1b) is satisfied for an arbitrary envelope $\tilde{\varphi}_w(y)$ of the additional potential ($\varphi_w = \tilde{\varphi}_w(y) e^{ik_z z}$), whereas Eq. (1a) is satisfied with $\mathbf{E} = E_y \hat{\mathbf{y}}$ provided:

$$\frac{\partial E_y}{\partial y} = \beta_p^2 \varphi_w. \tag{6}$$

Thus, provided the envelope $\tilde{\varphi}_w(y)$ is chosen such that $\int_{-\infty}^{+\infty} \tilde{\varphi}_w(u) du = 0$ (e.g. any odd function localized in the vicinity of the origin), it follows that the corresponding wave beam, characterized by the envelope $\tilde{E}_y = \beta_p^2 \int_{-\infty}^{y} \tilde{\varphi}_w(u) du$, is also localized in space and propagates along the z-direction with no diffraction, despite the fact that host material is assumed linear! Thus, one can say that the PEC wire medium supports non-diffracting waves even in the limit of a linear response. This property is a consequence of the diffraction-less nature of TEM waves in PEC wire media and of the fact that the effective medium is characterized by extreme anisotropy, as widely discussed in the literature [13, 14, 21].

## IV. Numerical Results

As a first example of the application of the theory of the previous section, in Fig. 3 we plot the normalized field profiles associated with a spatial soliton with $k_z = 0.990 k_h$ at $\lambda_0$=1550 nm in an array of silver nanowires embedded in a self-defocusing host medium. It is supposed that the radius of the nanowires is $r_w = 20 nm$ and that the period is $a = 200 nm$. The effect of dielectric and metallic loss is neglected here, and will be discussed in details in the next section. It can be checked that the electric field and additional potential profiles associated with a given spatial-soliton depend on the specific value of the nonlinear parameter ($\alpha < 0$) as $1/\sqrt{|\alpha|}$. Due to this reason, the field profiles in Fig. 3 are given in normalized unities. Without loss of generality, the permittivity of the host medium for weak field amplitudes is taken equal to the permittivity of vacuum $\varepsilon_h^0 = 1.0\varepsilon_0$. The results remain qualitatively similar if one considers the more realistic situation $\varepsilon_h^0 > \varepsilon_0$. As seen in Fig. 3, the spatial-soliton is a q-TEM beam with $|E_z| \ll |E_t|$ and $|\varphi_w/a| \ll |E_t|$. The dot-dashed black lines in Fig. 3 represent the field profiles obtained based on the paraxial approximation described in Sect. III, whereas the solid blue lines represent the field profiles obtained using the more rigorous (non-paraxial) theory of Appendix A. The two approaches yield nearly coincident results, consistent with the fact that the beam is a q-TEM wave. In particular, it is seen in the lower-right panel of Fig. 3 that $n_{ef,h}^2 \approx n_w^2/\zeta_w$, consistent with Eq. (2). Note that in the region where the field intensity is stronger one has $n_{ef,h}^2 < 1$, because the dielectric host is a self-defocusing Kerr medium. The envelope of $E_y$ has even parity, whereas the envelopes of $E_z$ and $\varphi_w$ have odd parity with respect to the coordinate $y$.

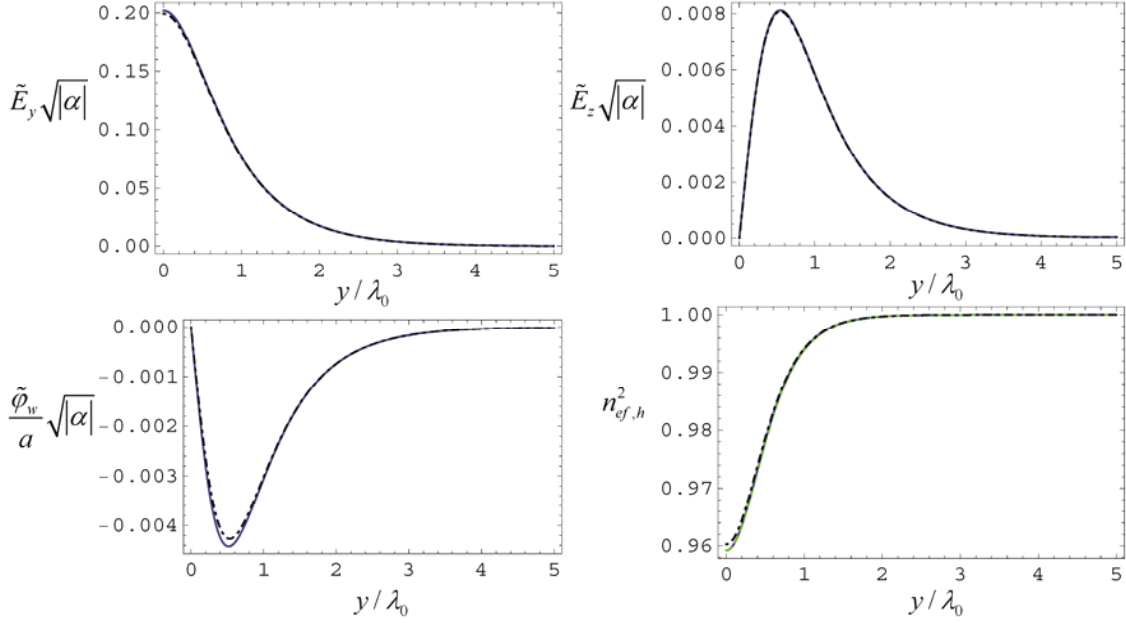

Fig. 3. (Color online) Normalized field profiles ($E_y = \tilde{E}_y(y)e^{ik_z z}$, $E_z = i\tilde{E}_z(y)e^{ik_z z}$ and $\varphi_w = \tilde{\varphi}_w(y)e^{ik_z z}$) for a spatial-soliton with $k_z = 0.990 k_h$ at $\lambda_0$=1550 nm in a nanowire material formed by silver wires in a dielectric background with $\varepsilon_h^0 = 1.0\varepsilon_0$, $r_w = 20nm$, $r_w/a = 0.1$. Solid blue lines: profiles calculated based on the non-paraxial approximation. Dot-dashed black lines: profiles calculated based on the paraxial approximation. The dashed green line (virtually coincident with the solid blue line) in the lower-right panel represents $n_w^2/\zeta_w$.

It can be checked that for the example of Fig. 3 the half-power beamwidth (which is roughly determined by the envelope of $E_y$) is such that $W = 1.06\lambda_0$ and thus it is of the order of the free-space wavelength. As could be expected, the beamwidth depends on the strength of the nonlinear effects, and for greater field intensities (smaller values of $k_z$) $W$ becomes increasingly smaller, and may even become subwavelength. This is illustrated in Fig. 4 where the half-power beamwidth is depicted as a function of the required value of $n_{ef,h}^2$ at the center of the beam. The results of Fig. 4 were obtained based on the non-paraxial approximation. Unfortunately, the condition to have a subwavelength solitons may require unrealistically strong nonlinear effects, associated with a significant depression of $n_{ef,h}^2$ (at least 5%) as compared to the case of a linear response.

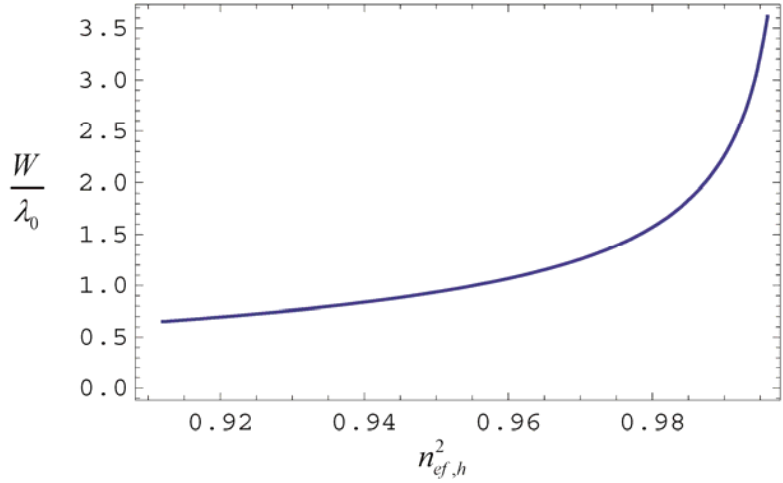

Fig. 4. (Color online) Normalized half-power-beamwidth ($W$) of the spatial-solitons as a function of $n_{ef,h}^2$ at $\lambda_0$=1550 nm in a nanowire material formed by silver wires in a dielectric background with $\varepsilon_h^0 = 1.0\varepsilon_0$, $r_w = 20nm$, $r_w/a = 0.1$.

To understand how the specific geometry of the nanowire material affects the profile of the spatial-solitons, we computed the solitons associated $k_z = 0.993k_h$ in a material formed by an array of silver nanowires embedded in a self-defocusing dielectric, considering different frequencies of operation, and different nanowires radii and period. The results of Fig. 5 indicate that for a fixed value of $k_z/k_h$ the electric field amplitude required to have the formation of a spatial-soliton is nearly independent of $r_w, a, \lambda_0$.

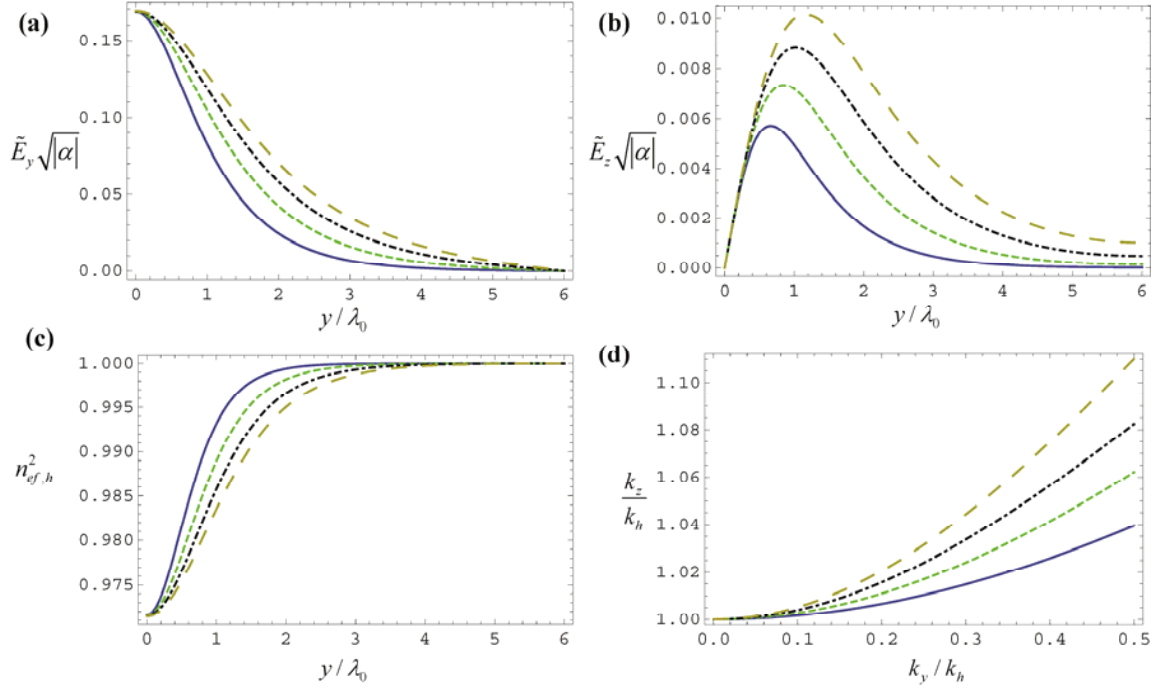

Fig. 5. (Color online) (a) and (b) Normalized field profiles ( $E_y = \tilde{E}_y(y)e^{ik_z z}$, $E_z = i\tilde{E}_z(y)e^{ik_z z}$ ) for a spatial-soliton with $k_z = 0.993 k_h$ in a material formed by silver nanowires in a dielectric background with $\varepsilon_h^0 = 1.0\varepsilon_0$. Solid blue lines: $r_w = 20nm$, $a = 200nm$, $\lambda_0 = 1550nm$; Dashed green lines: $r_w = 20nm$, $a = 200nm$, $\lambda_0 = 1300nm$; Dot-dashed black lines: $r_w = 20nm$, $a = 267nm$, $\lambda_0 = 1550nm$; Long dashed dark yellow lines: $r_w = 14nm$, $a = 200nm$, $\lambda_0 = 1550nm$; (c) profile of the index of refraction associated with the spatial-soliton. (d) Dispersion of the photonic states associated with the extraordinary (q-TEM) wave in case of a linear host with $n_{ef,h}^2 = n_w^2 / \zeta_w = 1.0$.

In addition, Fig. 5d shows that the characteristic spatial width of the soliton depends directly on the degree of hyperbolicity of the dispersion curve of the photonic states in the linear case. Specifically, if for a specific combination of the parameters $r_w, a, \lambda_0$ the dispersion curve of the photonic states becomes more hyperbolic the beamwidth of solitary wave becomes larger (see Figs. 5a and 5d). This result is consistent with the property discussed in Sect. III that in the limit of flat dispersion contours (case of a PEC metal) the nanowire material can support non-diffracting waves with an arbitrary transverse beamwidth length, even in the limit of vanishingly small nonlinear effects. Because of the plasma-type electrical response of the nanowires, the isofrequency contours of the photonic modes become more hyperbolic for shorter wavelengths and very dilute systems ($r_w/a$ very small). As a consequence in these scenarios the beamwidth of the spatial-solitons tends to increase, as illustrated in Fig. 5a.

## V. Effect of Loss

In the previous sections, the impact of dielectric and metal loss was not taken into account. In order to characterize the characteristic propagation length of the spatial-solitons in more realistic scenarios, next we investigate the spatial evolution of the solitary wave in the presence of material loss.

To this end, we consider again Eqs. (3) with $\mathbf{j}_{ext}=0$, $E_x=0$ and $\partial_x=0$. It is simple to verify that $\nabla\times\mathbf{E}=F_x\hat{\mathbf{x}}$ with $F_x=\dfrac{\partial E_z}{\partial y}-\dfrac{\partial E_y}{\partial z}$. Evidently, $F_x$ is proportional to the x-component of the macroscopic magnetic field: $F_x=i\omega\mu_0 H_x$. It is simple to check that Eqs. (3) can be rewritten as:

$$\frac{\partial F_x}{\partial z}=k_h^2 n_{ef,h}^2 E_y, \tag{7a}$$

$$\frac{\partial \varphi_w}{\partial z}=\frac{\zeta_w}{\beta_p^2}\left[-\frac{\partial F_x}{\partial y}-\left(k_h^2 n_{ef,h}^2-\frac{\beta_p^2}{\zeta_w}\right)E_z\right]. \tag{7b}$$

$$\frac{\partial E_z}{\partial z}=\left(k_h^2\zeta_w n_{ef,h}^2+\frac{\partial^2}{\partial z^2}\right)\varphi_w. \tag{7c}$$

$$\frac{\partial E_y}{\partial z}=\frac{\partial E_z}{\partial y}-F_x. \tag{7d}$$

Since the dominant component of the solitary wave is $E_y$, we can assume that $\dfrac{\partial}{\partial z} \to ik_{z0}$ when the operator $\partial_z$ acts over $\varphi_w$ and $E_z$. Hence, Eqs. (7b) and (7c) yield:

$$ik_{z0}\varphi_w = \frac{\zeta_w}{\beta_p^2}\left[-\frac{\partial F_x}{\partial y} - \left(k_h^2 n_{ef,h}^2 - \frac{\beta_p^2}{\zeta_w}\right)E_z\right], \tag{8a}$$

$$ik_{z0}E_z = \left(k_h^2 \zeta_w n_{ef,h}^2 - k_{z0}^2\right)\varphi_w. \tag{8b}$$

Substituting Eq. (8a) into (8b) it is found that:

$$\left(1 - \frac{\beta_p^2}{k_h^2 \zeta_w n_{ef,h}^2 - k_{z0}^2}\right) k_h^2 n_{ef,h}^2 E_z = -\frac{\partial F_x}{\partial y}. \tag{9}$$

Substituting this result into Eq. (7d) it follows that:

$$\frac{\partial E_y}{\partial z} = -\frac{\partial}{\partial y}\left[\left(1 - \frac{\beta_p^2}{k_h^2 \zeta_w n_{ef,h}^2 - k_{z0}^2}\right)^{-1} \frac{1}{k_h^2 n_{ef,h}^2} \frac{\partial F_x}{\partial y}\right] - F_x. \tag{10}$$

Because $n_{ef,h}^2$ is expected to vary relatively slowly in space, it is possible to neglect its derivatives with respect to $y$ and $z$. Differentiating both members of Eq. (10) with respect to $z$ and using Eq. (7a), it follows after some simplifications that:

$$\frac{\partial^2 E_y}{\partial z^2} + k_h^2 n_{ef,h}^2 E_y = -\left(1 - \frac{\beta_p^2}{k_h^2 \zeta_w n_{ef,h}^2 - k_{z0}^2}\right)^{-1} \frac{\partial^2 E_y}{\partial y^2}. \tag{11}$$

Writing $E_y = \tilde{E}_y(y,z) e^{ik_{z0}z}$ and using the standard paraxial approximation $\dfrac{\partial^2}{\partial z^2} \to 2ik_{z0}\dfrac{\partial}{\partial z} - k_{z0}^2$, one finds an equation that enables calculating the spatial evolution of the field envelope $\tilde{E}_y(y,z)$ along the direction of propagation:

$$2ik_{z0}\frac{\partial \tilde{E}_y}{\partial z} = -\left(1 - \frac{\beta_p^2}{k_h^2 \zeta_w n_{ef,h}^2 - k_{z0}^2}\right)^{-1} \frac{\partial^2 \tilde{E}_y}{\partial y^2} - \left(k_h^2 n_{ef,h}^2 - k_{z0}^2\right)\tilde{E}_y. \tag{12}$$

Specifically the spatial evolution is determined assuming that $\tilde{E}_y(y, z=0)$ is the profile of the spatial-soliton associated with $k_z = k_{z0}$ in the lossless case, calculated as explained in Sect. III. Then, the evolution of $\tilde{E}_y(y,z)$ is determined by solving Eq. (12) including the effect of metal loss in the parameter $\zeta_w$ and the effect of dielectric loss in $k_h = \omega\sqrt{\varepsilon_h^0 \mu_0}$, and using the boundary conditions $\tilde{E}_y(\pm y_{max}, z) = 0$, being $y_{max} \gg W$ and $W$ the beamwidth of the spatial-soliton. In our calculations, the effect of loss in silver was modeled by assuming that the collision frequency associated with the Drude model is 4.35THz [20].

We calculated the amplitude of the field envelope profile for the case of a spatial soliton associated with $k_z = 0.990 k_h$ at $\lambda_0$=1550 nm, which corresponds to the same example as in Fig. 3. The normalized calculated $|\tilde{E}_y|$ after a propagating distance of $12\lambda_0$ is shown in Fig. 6a, for the lossless case (solid blue line) and several lossy cases. As shown, by the (black) dashed curve the effect of metal loss is relatively weak and the reduction in the soliton amplitude is almost insignificant. This is explained in part by the relatively weak loss of silver in the infrared regime. However, the main reason for the weak sensitivity to loss in the metal is the fact that the solitary wave is a quasi-TEM beam (see Fig. 3), and consequently it interacts relatively weakly with the metallic wires, being the electric field mostly concentrated in the dielectric host. Consistent with this observation, it is seen in Fig. 6a (dark yellow long dashed curve) that the amplitude of the soliton is notably affected by the dielectric loss in the self-defocusing host dielectric. The green dashed curve in Fig. 6a, calculated for the case were metal and dielectric loss are both considered, further supports that the impact of metal loss is negligible as compared to dielectric loss.

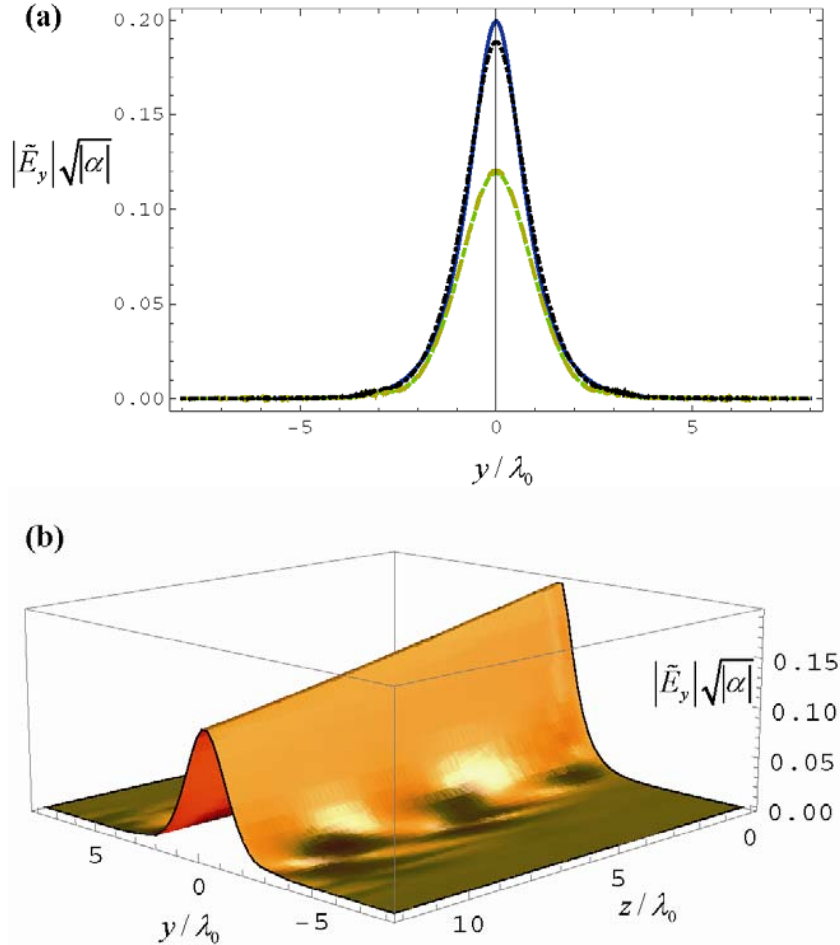

Fig. 6. (Color online) Normalized field profile of a spatial-soliton with $k_z = 0.990 k_h$ at $\lambda_0$=1550 nm in a nanowire material formed by silver wires in a dielectric background with $\operatorname{Re}\{\varepsilon_h^0\} = 1.0\varepsilon_0$, $r_w = 20 nm$, $r_w/a = 0.1$, after a propagation distance of $12\lambda_0$. (a) Solid blue line: loss effect is neglected. Green dashed line: effect loss in the silver nanowires and host medium is taken into account. Black dashed line: only loss in the metal is taken into account. Dark yellow long dashed line (practically coincident with the green line): only loss in the host dielectric is taken into account. (b) Profile of the spatial-soliton when loss in both the dielectric and silver are considered. The host medium is modeled as a lossy dielectric with loss tangent $\tan\delta = 0.01$.

Figure 6b reports the detailed spatial evolution of the solitary wave when loss in both the dielectric and silver is considered.

Our numerical simulations, based on the spatial evolution of Eq. (12), also seem to indicate that the profile of the solitary waves is stable to perturbations. It is interesting to note that the considered spatial-solitons cannot, as a result of some perturbation, be collapsed into an increasingly narrower beam because at a scale length of the order of the period $a$ the effective medium approximation ceases to work. In particular, a beam more localized than the period of the nanowire material will effectively see the "self-defocusing" host. Thus, the granularity of the lattice prevents the concentration of the solitary beam in a scale length smaller than the period $a$.

## VI. Conclusion

Based on an effective medium approximation, we characterized the spatial-solitons supported by metallic nanowire arrays embedded in a Kerr-type medium. We studied the physical requirements for the formation of spatial-solitons in nanowires structures. Taking into account the dynamics of the isofrequency surfaces of the photonic states in the nanowire material with respect to intensity of the electromagnetic field, it was shown that within the continuous medium approximation spatial solitons are supported only when the host dielectric is a self-defocusing material. It was demonstrated that for sufficiently strong electric field amplitude the spatial-solitons may become subwavelength, and that the main decay channel for the solitons is related to the loss in the dielectric, being the metallic loss comparatively less relevant. The developed theory provides simple means to characterize and analyze the formation and propagation of optical lattice solitons in metallic nanowire structures.

**Acknowledgments:**

The author gratefully acknowledges fruitful discussions with Fabio Biancalana and Andrea Marini. This work is supported in part by Fundação para Ciência e a Tecnologia under project PTDC/EEI-TEL/2764/2012.

# Appendix A

Here, the formalism used to determine the spatial solitons is developed. First, we consider the more general case (non-paraxial approach) wherein only the contribution of $|E_z|^2$ to $n_w^2$ and $n_{ef,h}^2$ is neglected, so that:

$$n_w^2 \approx \zeta_w \left[ 1 + \alpha \left( \mathbf{E}_t^* \cdot \mathbf{E}_t + \tilde{B} \beta_p^2 \varphi_w \varphi_w^* \right) \right], \tag{A1a}$$

$$n_{ef,h}^2 \approx 1 + \alpha \left( \mathbf{E}_t^* \cdot \mathbf{E}_t + \beta_p^2 \varphi_w \varphi_w^* \right). \tag{A1b}$$

We want to solve the nonlinear system (1) based on the above approximate formulas for $n_w^2$ and $n_{ef,h}^2$, and using the exact formula of $\tilde{\mathbf{Y}}$. We are interested in solitons such that the dependence on $z$ of $(\mathbf{E}, \varphi_w)$ is of the form $e^{ik_z z}$, $E_x = 0$ and $\partial_x = 0$ (the wave propagation and the electric field are in the $yoz$ plane). In this case, Eq. (1a) with $\mathbf{j}_{ext} = 0$ can be spelled out as follows:

$$ik_z \frac{\partial E_z}{\partial y} = -\left( k_z^2 - k_h^2 n_{ef,h}^2 \right) E_y + \beta_p^2 k_h^2 \tilde{Y}_y \varphi_w, \tag{A2a}$$

$$-\frac{\partial^2 E_z}{\partial y^2} + ik_z \frac{\partial E_y}{\partial y} = \left( k_h^2 n_{ef,h}^2 - \frac{\beta_p^2}{\zeta_w} \right) E_z + \frac{\beta_p^2}{\zeta_w} ik_z \varphi_w. \tag{A2b}$$

with $\tilde{Y}_y = \frac{\alpha}{2}\left(\varphi_w E_y^* + \varphi_w^* E_y\right)$. On the other hand, using Eq. (1b) it is possible to obtain an explicit expression for $E_z$ in terms of $\mathbf{E}_t$ and $\varphi_w$:

$$E_z = \left[\left(k_h^2 n_w^2 - k_z^2\right)\varphi_w + \zeta_w k_h^2 \tilde{\mathbf{Y}} \cdot \mathbf{E}_t\right]/(ik_z). \tag{A3}$$

It is proven in Appendix B that by substituting this result into Eqs. (A2) it is possible to eliminate $E_z$, and obtain a nonlinear system for $E_y$ and $\varphi_w$. We are interested in spatial solitons such that $E_y$ and $\varphi_w$ are in phase. Specifically, it is assumed that $E_y = \tilde{E}_y(y)e^{ik_z z}$ and $\varphi_w = \tilde{\varphi}_w(y)e^{ik_z z}$ with the envelopes $\tilde{E}_y$ and $\tilde{\varphi}_w$ real valued. In these circumstances, $E_y$ and $\varphi_w$ satisfy the following first order partial nonlinear system (see Appendix B),

$$\mathbf{A} \cdot \mathbf{y} = \mathbf{b}. \tag{A4}$$

where $\mathbf{y} = \left(\dfrac{\partial \varphi_w}{\partial y} \quad \dfrac{\partial E_y}{\partial y}\right)^T$ and the 2×2 matrix $\mathbf{A} = \left[a_{i,j}\right]$ has the elements:

$$a_{11} = k_h^2 n_w^2 - k_z^2 + \alpha \zeta_w k_h^2 E_y E_y^* + 2\alpha \zeta_w k_h^2 \tilde{B} \beta_p^2 \varphi_w \varphi_w^*, \tag{A5a}$$

$$a_{12} = \zeta_w k_h^2 \tilde{Y}_y + 3\alpha \zeta_w k_h^2 E_y \varphi_w^*, \tag{A5b}$$

$$a_{21} = -\left(\beta_p^2 \tilde{Y}_y + 3\alpha \beta_p^2 E_y \varphi_w^*\right), \tag{A5c}$$

$$a_{22} = -\left(n_{ef,h}^2 + 2\alpha E_y E_y^* + \alpha \beta_p^2 \varphi_w \varphi_w^*\right). \tag{A5d}$$

The vector $\mathbf{b}$ in the right-hand side of Eq. (A4) is given by:

$$\mathbf{b} = \begin{pmatrix} -\left(k_z^2 - k_h^2 n_{ef,h}^2\right)E_y + \beta_p^2 k_h^2 \tilde{Y}_y \varphi_w \\ \left(k_h^2 n_{ef,h}^2 \zeta_w - \beta_p^2\right)\tilde{Y}_y E_y + \left[\left(k_h^2 n_w^2 - k_z^2\right)n_{ef,h}^2 - \dfrac{\beta_p^2}{\zeta_w}n_w^2\right]\varphi_w \end{pmatrix}. \tag{A6}$$

The reason why we look for spatial solitons such that $E_y$ and $\varphi_w$ are in phase is that in such circumstances it follows from Eq. (A3) that $E_y$ and $E_z$ are in quadrature in the limit of negligible material loss. Note that when $E_y$ and $\varphi_w$ are in phase and when loss is negligible $n_w^2$, $\zeta_w$, and $\tilde{Y}_y$ are real-valued. For a plane wave natural mode in the uniaxial wire medium, it is easy to prove that in the linear case ($\alpha = 0$) and in the limit of negligible loss the phase difference between $E_y$ and $E_z$ (with $E_x = 0$) is the same as the phase difference between $k_y$ and $k_z$, being $\mathbf{k} = (0, k_y, k_z)$ the wave vector of the plane wave [13, 14]. Thus, when $E_y$ and $E_z$ are in quadrature the wave vector components $k_y$ and $k_z$ also are. In particular, when $k_z$ is real valued the natural mode has an exponential-type variation along the $y$ direction. This suggests that solutions of Eqs. (A2) with $E_y$ and $E_z$ in quadrature and a variation along $z$ of the form $e^{ik_z z}$ with $k_z$ is real valued may be associated with waves that decay exponentially when $y \to \pm\infty$ and propagate along $z$, which is the desired behavior for the spatial solitons.

In the paraxial approximation only the contributions to the nonlinear dynamics arising from terms of the form $E_y E_y^*$ should be retained. Hence, within this approximation, one should set equal to zero terms of form $\tilde{Y}_y$, $\varphi_w \varphi_w^*$ and $E_y \varphi_w^*$ in Eqs. (A5)-(A6). It can be readily checked that this yields the nonlinear system of Eqs. (3a)-(3b).

## Appendix B

To obtain the nonlinear differential system (A4), first we calculate $\dfrac{\partial^2 E_z}{\partial y^2} = \dfrac{\partial}{\partial y}\dfrac{\partial E_z}{\partial y}$ in the left-hand side of Eq. (A2b) with the help of the explicit formula for $\dfrac{\partial E_z}{\partial y}$ given by Eq. (A2a). This gives us:

$$\frac{\partial}{\partial y}\left[k_h^2 n_{ef,h}^2 E_y + \beta_p^2 k_h^2 \tilde{Y}_y \varphi_w\right] = (-ik_z)\left(k_h^2 n_{ef,h}^2 - \frac{\beta_p^2}{\zeta_w}\right)E_z + \frac{\beta_p^2}{\zeta_w}k_z^2 \varphi_w. \tag{B1}$$

By substituting Eq. (A3) into Eqs. (A2a) and (B1), it is found after straightforward simplifications that:

$$\frac{\partial}{\partial y}\left[\left(k_h^2 n_w^2 - k_z^2\right)\varphi_w + \zeta_w k_h^2 \tilde{Y}_y E_y\right] = b_1, \tag{B2a}$$

$$-\frac{\partial}{\partial y}\left[n_{ef,h}^2 E_y + \beta_p^2 \tilde{Y}_y \varphi_w\right] = b_2. \tag{B2b}$$

where $b_1$ and $b_2$ are such that $\mathbf{b} = (b_1 \; b_2)^T$ is given by Eq. (A6). Next, we calculate explicitly the derivatives in the left-hand side of Eqs. (B2) with the help of $\tilde{Y}_y = \frac{\alpha}{2}\left(\varphi_w E_y^* + \varphi_w^* E_y\right)$ and of the formulas for $n_w^2$ and $n_{ef,h}^2$ [Eqs. (A1a)-(A1b)]. This yields a nonlinear system that can be written in the matrix form,

$$\mathbf{M} \cdot \mathbf{x} = \mathbf{b} \tag{B3}$$

with $\mathbf{x} = \left(\dfrac{\partial \varphi_w}{\partial y} \;\; \dfrac{\partial \varphi_w^*}{\partial y} \;\; \dfrac{\partial E_y}{\partial y} \;\; \dfrac{\partial E_y^*}{\partial y}\right)^T$. The 2×4 matrix $\mathbf{M} = \left[m_{i,j}\right]$ has the elements:

$$m_{11} = k_h^2 n_w^2 - k_z^2 + \frac{\alpha}{2}\zeta_w k_h^2 E_y E_y^* + \alpha \zeta_w k_h^2 \tilde{B} \beta_p^2 \varphi_w \varphi_w^* \tag{B4a}$$

$$m_{12} = \alpha \zeta_w k_h^2 \tilde{B} \beta_p^2 \varphi_w^2 + \frac{\alpha}{2}\zeta_w k_h^2 E_y^2 \tag{B4b}$$

$$m_{13} = \zeta_w k_h^2 \tilde{Y}_y + \frac{\alpha}{2}\zeta_w k_h^2 E_y \varphi_w^* + \alpha \zeta_w k_h^2 \varphi_w E_y^* \tag{B4c}$$

$$m_{14} = \frac{3}{2}\alpha \zeta_w k_h^2 E_y \varphi_w \tag{B4d}$$

$$m_{21} = -\left(\beta_p^2 \tilde{Y}_y + \alpha \beta_p^2 E_y \varphi_w^* + \frac{\alpha}{2}\beta_p^2 \varphi_w E_y^*\right) \tag{B4e}$$

$$m_{22} = -\frac{3}{2}\alpha \beta_p^2 \varphi_w E_y \tag{B4f}$$

$$m_{23} = -\left( n_{ef,h}^2 + \alpha E_y E_y^* + \frac{\alpha}{2} \beta_p^2 \varphi_w \varphi_w^* \right) \qquad (B4g)$$

$$m_{24} = -\alpha \left( E_y^2 + \frac{1}{2} \beta_p^2 \varphi_w^2 \right) \qquad (B4h)$$

In case $E_y$ and $\varphi_w$ are assumed to be in phase, it is straightforward to check that the system (B3) reduces to Eq. (A4).